# Shaping the modal confinement in silicon nanophotonic waveguides through dual-metamaterial engineering

*Thi Thuy Duong Dinh[1,*], Xavier Le Roux[1], Jianhao Zhang[1], Miguel Montesinos-Ballester[1], Christian Lafforgue[1], Daniel Benedikovic[2], Paben Cheben[3,4], Eric Cassan[1], Delphine Marris-Morini[1], Laurent Vivien[1], Carlos Alonso-Ramos[1]*

[1]Universite Paris-Saclay, Univ. Paris-Sud, CNRS Centre de Nanosciences et de Nanotechnologies, 91120, Palaiseau, France
[2]Dept. Multimedia and Information-Communication Technologies, University of Zilina, 01026 Zilina, Slovakia
[3]National Research Council Canada, Ottawa K1A 0R6, Canada
[4]Center for Research in Photonics, University of Ottawa, Ottawa K1N6N5, Canada

Email Address: thi-thuy-duong.dinh@c2n.upsaclay.fr

**Flexible control of the modal confinement in silicon photonic waveguides is an appealing feature for many applications, including sensing and hybrid integration of active materials. In most cases, strip waveguides are the preferred solution to maximize the light interaction with the waveguide surroundings. However, the only two degrees of freedom in Si strip waveguides are the width and thickness, resulting in limited flexibility in evanescent field control. Here, we propose and demonstrate a new strategy that exploits metamaterial engineering of the waveguide core and cladding to control the index contrast in the vertical and horizontal directions, independently. The proposed dual-material geometry yields a substantially increased calculated overlap with the air (0.35) compared to the best-case scenario for a strip waveguide (0.3). To experimentally demonstrate the potential of this approach, we have implemented dual-metamaterial ring resonators, operating with the transverse-magnetic polarized mode in 220-nm-thick waveguides with air as upper-cladding. Micro-ring resonators implemented with strip and dual-metamaterial waveguides exhibit the same measured quality factors, near 30,000. Having similar measured quality factors and better calculated external confinement factors than strip waveguides, the proposed dual-metamaterial geometry stands as a promising approach to control modal confinement in silicon waveguides.**

## 1.Introduction

Driven by the impressive technology development in recent years, silicon photonics is expanding its frontiers towards new applications beyond telecom and Datacom, with a particular interest in chemical and biological sensing [1, 2, 3]. Concurrently, novel materials are being considered for hybrid integration with silicon photonics [4, 5]. Flexible control of the modal confinement in silicon nanophotonic waveguides is a particularly appealing feature that could be instrumental for the aforementioned applications that harness the interaction of light with the waveguide surroundings.

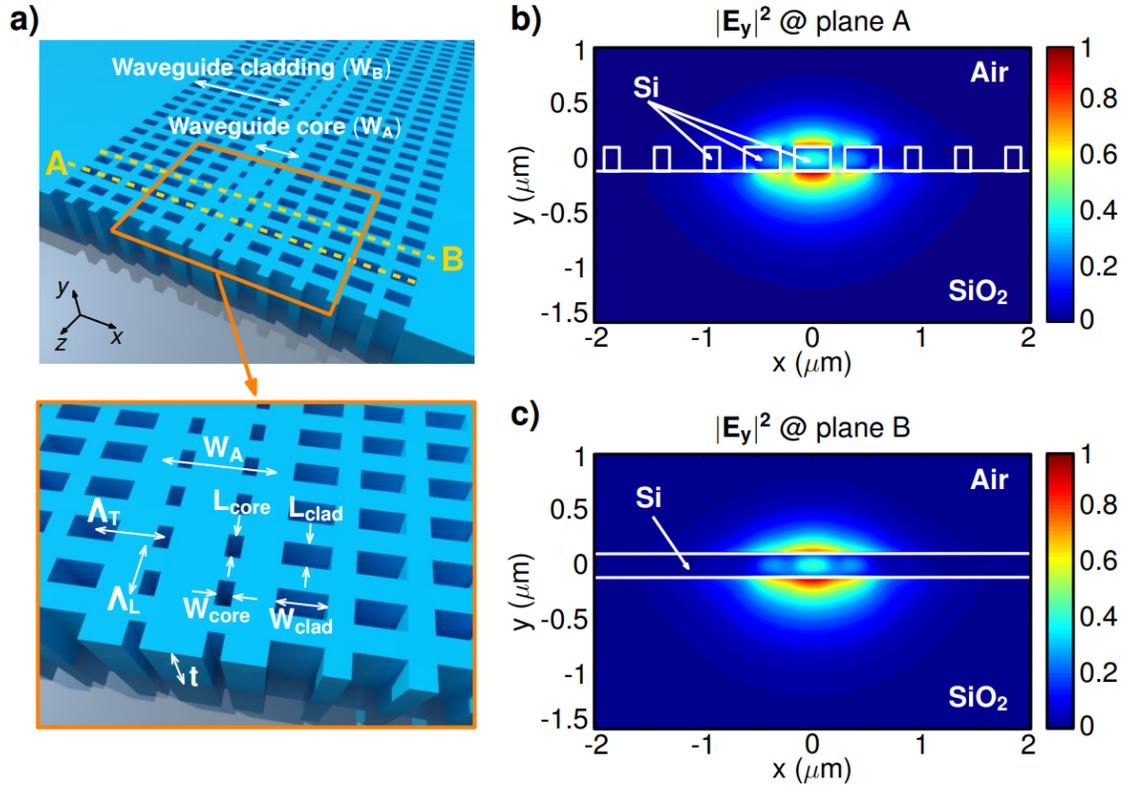

Figure 1: a) Schematic of the dual metamaterial waveguide. Electric field distribution of the fundamental TM mode b) at the plane A and c) at the plane B, defined in panel a). Wavelength is 1550 nm, and the geometrical parameters are: $\Lambda_L$ = 200 nm, $\Lambda_T$ = 450 nm, $L_{core} = L_{clad}$ = 100 nm, $W_{core}$ = 100 nm and $W_{clad}$ = 300 nm. The number of holes in the core and cladding are $N_{core}$ = 2, and $N_{clad}$ = 16, respectively.

Several geometries have been considered to control the field confinement, aiming at maximizing the interaction of light with the waveguide surroundings through the optical evanescent field. For instance, slot waveguides [6], and slot photonic crystal waveguides [7] operating with transverse-electric (TE) polarization can yield remarkable evanescent fields. However, the optical modes of these waveguides have a strong interaction with the vertical walls, where the sidewall roughness is, resulting in comparatively high propagation loss that compromises their feasibility. On the other hand, subwavelength metamaterial engineering has been identified as a promising tool to shape modal confinement in silicon photonic waveguides [8]. Subwavelength grating (SWG) metamaterial waveguides rely on periodic silicon patterning, with a structural period shorter than half the wavelength, to synthesize refractive index that can have any value between those of the silicon and the cladding [9, 10, 11]. Unlike photonic crystals that rely on resonant light confinement, SWG waveguides operate well below the bandgap, guiding light by metamaterial refractive index difference. This way, they provide low propagation loss and remarkably wide spectral bandwidth. In principle, the optical mode confinement can be shaped at will just by reducing the metamaterial refractive index. However, the minimum metamaterial index and maximum evanescent field that can be implemented in practice are limited by the loss due to leakage into the silicon substrate [12]. Most state-of-the-art SWG waveguides rely on metamaterial engineering of the core [8, 13] or the cladding [14, 15, 16, 17], but not both. In these geometries, the index of the cladding or the core is fixed, thereby limiting the flexibility in controlling the modal confinement in the vertical and horizontal directions, independently. In waveguides with a subwavelength metamaterial cladding only, the evanescent field in the horizontal dimension can be tuned by controlling refractive index contrast between the waveguide core and cladding, and by engineering cladding anisotropy [15, 16, 17]. However, the achievable evanescent field in the vertical dimension is limited by the tight confinement in the central silicon strip. On the other hand, in waveguides with a subwavelength metamaterial core, reducing the metamaterial index expands the mode in the vertical and horizontal directions, simultaneously. Thus, maximum evanescent field is severely limited by leakage into the silicon substrate due to excessive vertical expansion, while the latter can be reduced by increasing the waveguide width. Still, the maximum waveguide width is restricted by the single-mode operation. The leakage into the substrate is a particularly relevant limitation when air serves as a top-cladding material (superstrate), as it is the case in gas sensing applications [18]. The large asymmetry between the top and bottom cladding results in large loss due to leakage into the substrate that precludes the use of metamaterial waveguides with the nanostructured core. This is especially the case when considering the widely used silicon-on-insulator (SOI) technology with 220-nm-thick silicon guiding layer and wavelengths near 1.5 µm. This limitation could be alleviated by removing the buried oxide (BOX) layer to implement a silicon membrane with air as upper and lower cladding [14]. For instance, subwavelength membrane waveguides interleaving shifted silicon strips have been recently reported that implement different metamaterials in the core and the cladding [19]. However, the same silicon strips define the core and cladding geometry, limiting the flexibility in the design of the index contrast. In addition, the use of membrane waveguides complicates fabrication process, compromising the mechanical stability of the device. After all, the preferred solution for most sensing applications is the strip waveguide, operating with transverse-magnetic (TM) polarized mode that provides comparatively large evanescent field with a weaker interaction with sidewall roughness, thus yielding lower propagation loss. Nevertheless, for a given SOI technology, i.e. fixed thickness of Si guiding layer, the only degree of freedom available to control modal confinement in strip waveguides is the core width. This little design flexibility limits the achievable evanescent field and the interaction with the waveguide surroundings.

Here, we propose and experimentally demonstrate a new waveguide geometry that implements different subwavelength metamaterials in the core and the cladding (see Fig. 1). This strategy enables flexible control of the synthesized metamaterial indices in the core and the cladding, independently, thereby releasing new degrees of freedom to shape modal confinement. More specifically, the vertical index contrast is primarily governed by the metamaterial in the waveguide core, while the lateral index contrast is set by the difference between the metamaterial indices in the core and cladding. By reducing the lateral index contrast, we favor field expansion in the horizontal direction, while affording single-mode operation for comparatively wider waveguides. By increasing the waveguide width and enhancing the lateral expansion we substantially relax limitations in the core metamaterial index due to leakage into the silicon substrate. To demonstrate the potential of this approach, we have implemented dual-metamaterial waveguides in the SOI platform with a 220-nm-thick guiding silicon layer and air as a top cladding, operating with TM polarization near 1.5 µm wavelength. Our calculations show that the proposed waveguide ensures single mode operation for core widths as large as 900 nm, yet achieving better overlap with the surroundings than the TM modes of the strip waveguide. To validate the practical feasibility of the proposed concept, we have fabricated ring resonators with dual-metamaterial waveguides showing measured quality factors larger than 30,000 at critical coupling. For comparison, we fabricated and characterized strip ring resonators, that showed similar quality factors. These results prove the feasibility of metamaterial engineering in applications requiring air as top cladding, and open promising prospects for applications of the dual-metamaterial engineering for controlling the modal confinement in silicon nanophotonic waveguides.

## 2. Dual-metamaterial waveguide: operation principle and design

The proposed dual-metamaterial waveguide geometry, depicted in Fig. 1(a), comprises multiple holes that are placed periodically within the core and the cladding. All the holes in the lattice have the same longitudinal ($\Lambda_L$) and transversal ($\Lambda_T$) periods. In the proposed design, both periods are short enough to ensure subwavelength operation. The waveguide core has a width of $W_A$, comprising a series of holes with a length and width of $L_{core}$ and $W_{core}$, respectively. The holes in the cladding have a length of $L_{clad}$ and width of $W_{clad}$. In each longitudinal period, the waveguide comprises $N_{core}$ holes in the core and $N_{clad}$ holes at each side of the cladding. For the waveguide design, we have considered a silicon layer thickness of 220 nm and air as upper cladding. The transverse-electric (TE) polarized modes typically have a strong field near the vertical waveguide walls, affected by the sidewall roughness effect, which may result in increased propagation loss. To circumvent this limitation, we decided to work with the TM polarized mode. We have chosen longitudinal and transversal periods of $\Lambda_L$ = 200 nm and $\Lambda_T$ = 450 nm, respectively, that ensure suppression of the diffraction effects in the longitudinal [10] and transversal [20] directions, respectively. The number of holes in the core is $N_{core}$ = 2, and in the cladding, it is $N_{clad}$ = 16. The holes in the core and cladding have a length of $L_{core}$ = $L_{clad}$ = 100 nm, yielding a duty cycle of 50 % in the longitudinal direction. The holes in the core have a width of $W_{core}$ = 100 nm, while the holes in the cladding have a width of $W_{clad}$ = 300 nm. These widths have been chosen as a compromise between minimum feature size and index contrast between the core and cladding.

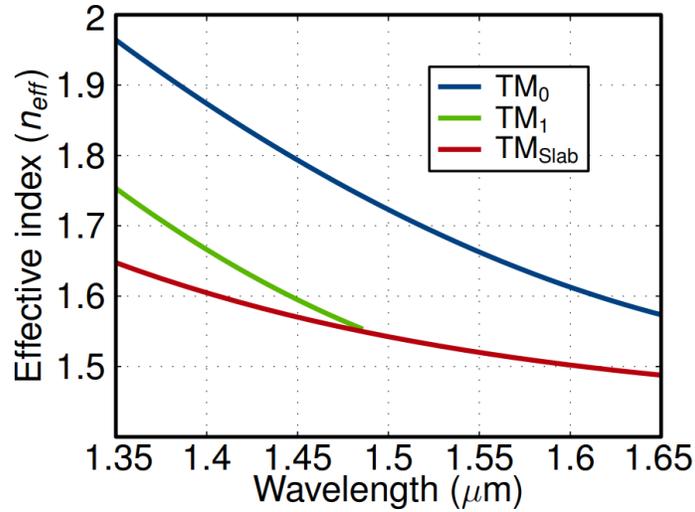

Figure 2: Effective index of the fundamental and first order TM modes of the dual-metamaterial waveguide, compared with the effective index of the metamaterial slab in the lateral cladding. These indices are calculated using 3D-FDTD simulations.

To study the properties of the proposed waveguide, we use three-dimensional finite-difference time domain (3D-FDTD) simulations [21]. We considered one unit-cell of the waveguide, with the longitudinal period $\Lambda_L$ = 200 nm. We applied periodic Bloch boundary conditions in the propagation direction and perfectly matched layer (PML) boundary conditions in the transversal directions. We considered the refractive index of $n_{air}$ = 1 for the upper cladding and of $n_{BOX}$ = 1.44 for the buried oxide. The electric field distribution for the TM mode, near 1550 nm wavelength, at the planes A and B, described in Fig.1(a), are shown in Figs. 1(b) and (c), respectively. The evanescent field concentrates at the top and bottom silicon surfaces, with a weak interaction with the vertical walls.

To illustrate the single-mode operation of the proposed waveguide, we calculate the effective index of its fundamental and first order TM modes as a function of the wavelength, comparing them with the effective index of the metamaterial cladding (see Fig. 2). We extract the effective indices from the analysis of the wavevector in 3D-FDTD calculations of the dual-metamaterial waveguide and a slab waveguide made with the metamaterial cladding. To be guided, the waveguide modes need to have an effective index higher than that of the metamaterial cladding. As shown in Fig. 2, the waveguide with a width of 900 nm (transversal period $\Lambda_T$ = 450 nm and $N_{core}$ = 2), yields single mode operation for wavelengths above 1.5 μm. The maximum waveguide width for single-mode operation for TM strip is ∼ 600 nm. We analyse the potential of the dual-metamaterial waveguide deconfine the optical mode. We decided to use a strip waveguide with TM polarization for comparison, as it is the preferred configuration for most applications requiring light interaction with the waveguide cladding. SWG waveguides and slot waveguides have been proposed as an alternative solution to strip waveguides maximizing the evanescent field, mainly for sensing applications where water serves as upper-cladding material (analyte) [18]. These geometries may achieve larger evanescent fields for TE-polarized modes than strip waveguides. However, they also exhibit stronger interaction with sidewall roughness, resulting in higher propagation loss that offsets the advantages of having a higher overlap with the cladding [3, 22]. SWG waveguides operating with TM polarization can have large evanescent fields with a weaker interaction with vertical walls, thus lower propagation loss [3]. However, their use in the case of air cladding is seriously hampered by leakage losses into the silicon substrate. As an illustrative example, a SWG waveguide with an infinite width, a silicon thickness of 220 nm, a silicon section length of 200 nm, gap section of length 100 nm, and air as top cladding yields a TM mode with an effective index of 1.47 at 1.5 µm wavelength, near the cut-off condition.

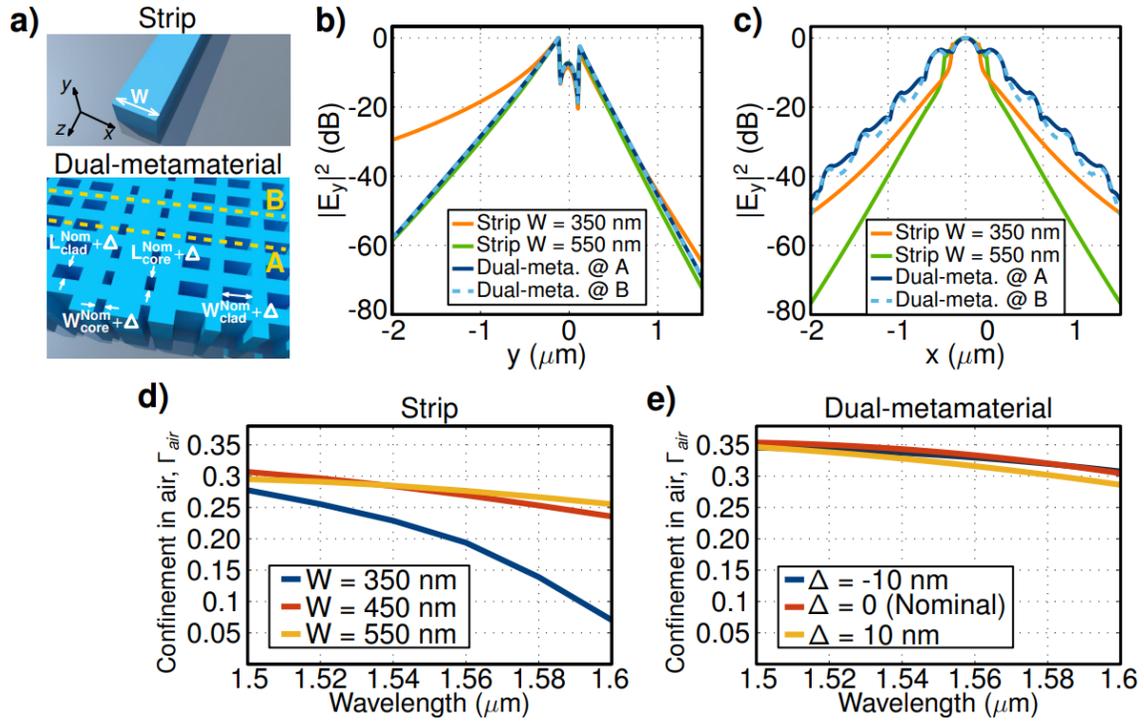

Figure 3: a) Schematic showing main geometrical parameters of the strip and dual-metamaterial waveguides. Comparison of the electric field distributions along the b) vertical and c) horizontal dimensions for the proposed dual-metamaterial waveguide and strip waveguide. Calculated external confinement factor in the air for d) strip and e) dual-metamaterial waveguides.

In Fig. 3(a), we define the main geometrical parameters considered for strip and dual-metamaterial waveguides. The performance of the strip waveguides is studied using 2D finite-difference eigenmode solver, while dual-metamaterial waveguides are analysed with 3D-FDTD simulations. In Figs. 3(b) and 3(c), we compare the field distributions for the TM modes of the strip and dual-metamaterial waveguides along with the horizontal and vertical directions, respectively. For the strip waveguide, we considered two waveguide widths of 350 nm and 550 nm. In the case of the 350 nm width, the optical mode is highly evanescent, having an effective index of 1.45 near the cut-off limit. The waveguide width of 550 nm is close to the limit of single-mode condition (600 nm), yielding a highly confined optical mode. The proposed dual-metamaterial waveguide exhibits a field profile along the vertical direction similar to that of the highly confined strip waveguide (W = 550 nm). Along the horizontal direction, the field profile of the dual-metamaterial waveguide resembles that of the highly evanescent strip waveguide (W = 350 nm). This particular field distribution illustrates the great flexibility in the engineering of field confinement provided by the proposed dual-metamaterial waveguide. An important parameter to assess the interaction of the mode with the waveguide surroundings is the external confinement factor in the upper cladding [22]. The higher the external confinement factor, the higher the interaction. The external confinement factor in the air, $\Gamma_{air}$, can be calculated as [3, 22]:

$$\Gamma = \frac{dn_{eff}}{dn_{clad}} = \frac{n_g}{n_{clad}} \frac{\int_{clad} \epsilon |\vec{E}|^2 d^2x}{\int \epsilon |\vec{E}|^2 d^2x}$$

where $n_{eff}$, $n_g$ and E are the effective index, group index, and electric field of the waveguide mode, respectively; $n_{air}$ = 1 is the refractive index of the air and $\epsilon$ is the permittivity. For the strip waveguide, we evaluate three waveguide widths (W) of 350 nm, 450 nm, and 550 nm, within the single-mode regime (maximum width of 600 nm). For the dual-metamaterial waveguide, we consider the nominal design ($L_{core}$ = $L_{clad}$ = 100 nm, $W_{core}$ = 100 nm, $W_{clad}$ = 300 nm, $N_{core}$ = 2) and geometrical variations due to fabrication imperfections. The length and width of the holes in the core and cladding are defined as $L_i = L_i^{Nom} + \Delta$ and $W_i = W_i^{Nom} + \Delta$. The subscript i denotes holes in the core (i = core) or cladding (i = clad), and $\Delta$ defines the deviation from the nominal dimension. As shown in Figs. 3(d) and 3(e) the dual-metamaterial waveguide yields a higher external confinement factor than the strip waveguide. The external field confinement of the dual-metamaterial waveguide is very robust against fabrication imperfections, with variations below 0.01 for errors as large as ±10 nm.

## 3 Experimental results

To evaluate the practical feasibility of the proposed approach we have fabricated and experimentally characterized multiple micro-ring resonators with dual-metamaterial waveguides. Micro-rings are widely used in applications exploiting resonant enhancement of the light-matter interaction like sensing [3] and frequency comb generation [23]. Achieving micro-ring resonators with high quality factors is key to maximize light-matter interaction enhancement due to resonant light recirculation. Achieving high quality factors requires not only low propagation loss, but also low bending loss and reduced coupling loss between the bus waveguide and the ring resonator. These parameters are essential for the successful exploitation of the proposed dual-metamaterial waveguides. For comparison, we have also fabricated and characterized strip ring resonators, operating with TM polarization.

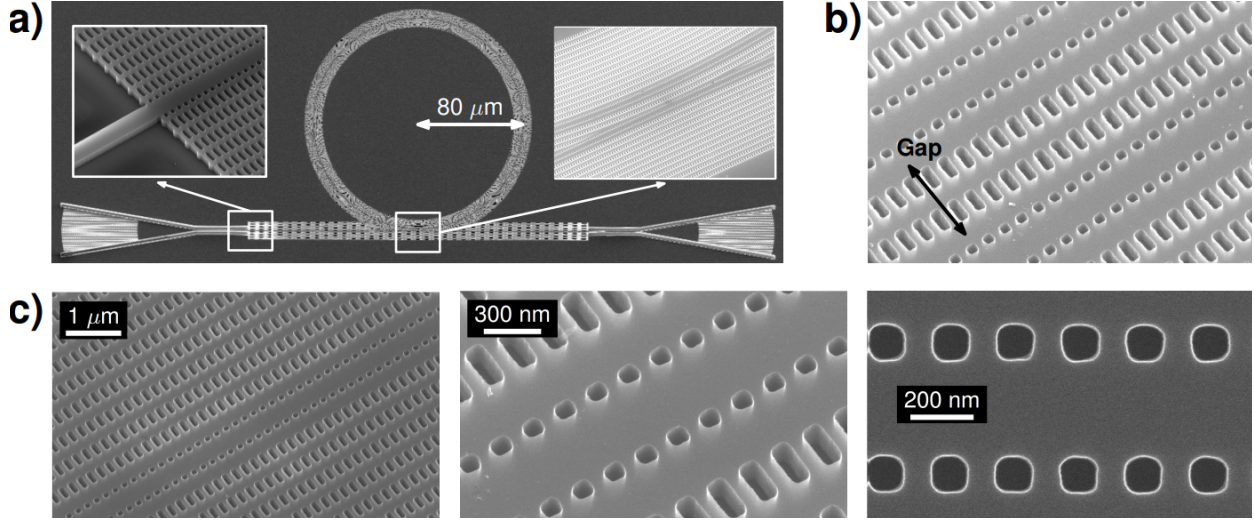

Figure 4: a) Schematic showing main geometrical parameters of the strip and dual-metamaterial waveguides. Comparison of the electric field distributions along the b) vertical and c) horizontal dimensions for the proposed dual-metamaterial waveguide and strip waveguide. Calculated external confinement factor in the air for d) strip and e) dual-metamaterial waveguides.

For the fabrication of the micro-ring resonators, we used SOI wafers with a 220-nm-thick guiding silicon layer and a 3-µm-thick BOX layer. The patterns were defined by electron beam lithography and transferred into silicon by a single-step reactive ion etching process. Figure 4 shows some scanning electron microscope (SEM) images of the dual-metamaterial ring resonators.

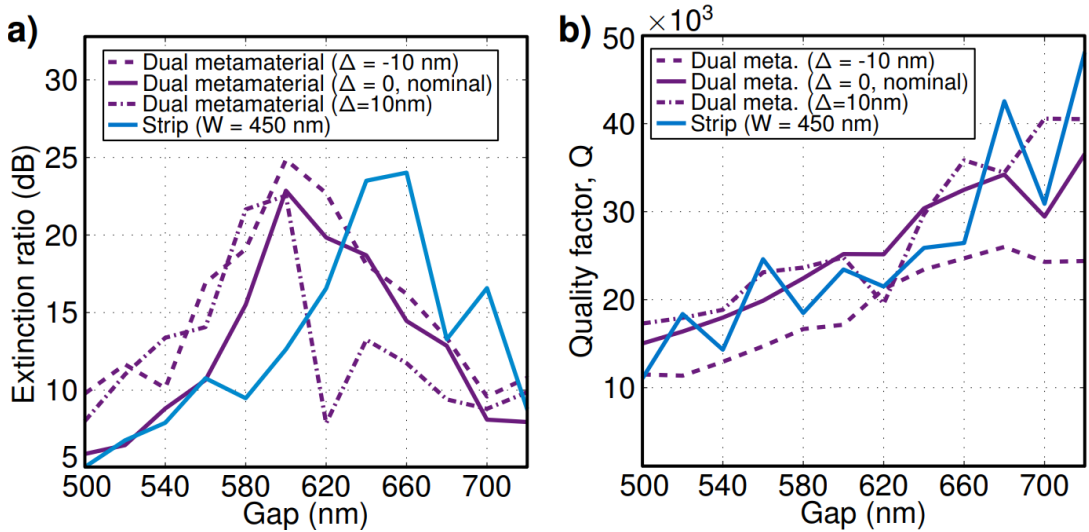

Figure 5: Comparison of measured a) extinction ratio and b) loaded quality factor near 1540 nm wavelength, as a function of the gap between the ring and the bus waveguide for three dual-metamaterial ring resonators designs (Δ = -10, 0, 10 nm) and strip ring, with waveguide width of 450 nm. The ring radius is 80 µm in all cases.

Light is injected and extracted from the chip using cleaved single-mode fibers (SMF-28) and fully-etched subwavelength-engineered fiber-chip grating couplers, optimized for TM polarization (pitch of 1.2 µm, duty cycle of 0.5 and coupling efficiency of -10 dB). The geometrical parameters of our nominal dual-metamaterial waveguide are $\Lambda_L$ = 200 nm, $\Lambda_T$ = 450 nm, $L_{core}$ = $L_{clad}$ = 100 nm, $W_{core}$ = 100 nm, $W_{clad}$ = 300 nm, $N_{core}$ = 2 and $N_{clad}$ = 16. Figure 4(a) (top-left) shows the transition implemented between the strip and the dual-metamaterial waveguide. In the beginning, the holes in the core have a length of 50 nm and width of 50 nm, while the holes in the cladding have a size of 150 nm × 400 nm. The size of the holes is changed adiabatically along 30 µm long taper to reach the values of the dual-metamaterial waveguide. This transition yields a measured insertion loss lower than 1.5 dB. For the strip resonators, the waveguide width is 450 nm. The radius is 80 µm for both the dual-metamaterial and the strip ring resonators. In both types of resonators, we varied the gap between 500 nm and 720 nm.

Figure 5 shows the comparison of measured extinction ratio, Fig. 5(a), and loaded quality factor, Fig.5(b), for a wavelength of 1540 nm, as a function of the gap between the ring and the bus waveguide for three dual-metamaterial ring resonators designs ($\Delta$ = -10, 0, 10 nm) and strip ring, with waveguide width of 450 nm. The three dual-metamaterial designs yield the critical coupling for a gap of 600 nm between the bus and the ring, with a maximum extinction ratio of ~ 25 dB comparable to the strip case. As illustrated in Fig. 5(b), the dual-metamaterial rings yield quality factors similar to those of the strip rings. In all cases, the quality factor increases monolithically with the gap width. This could be attributed to a reduction in the coupler loss for larger gaps. In Fig. 6 we compare the spectral variation of the extinction ratio and loaded quality factor of the nominal dual-metamaterial ring ($\Delta$ = 0 nm) and strip ring at critical coupling. For the dual-metamaterial ring the coupling gap is 600 nm while for the strip ring the gap is 660 nm. Figure 6(a) shows the transmittance spectrum of the dual-metamaterial ring, showing only one set of resonances, with a free-spectral range (FSR) near 1.4 nm at 1540 nm wavelength, and no sign of multimode behavior. From the measured FSR, we estimate a group index of 3.5, in good agreement with the calculated value of the fundamental TM mode (3.4). Figure 6(b) shows the transmittance spectrum for the strip ring resonator, achieving critical coupling near 1540 nm wavelength. The FSR for the strip ring resonator is ~1.4 near 1540 nm wavelength. Critical coupling is achieved for a considerably narrower spectral range in the case of the strip ring, which could be related to a more dispersive coupling between the ring and the bus waveguide. As shown in Fig. 6(c) both, strip and dual-metamaterial rings exhibit very similar quality factors in the measured wavelength range.

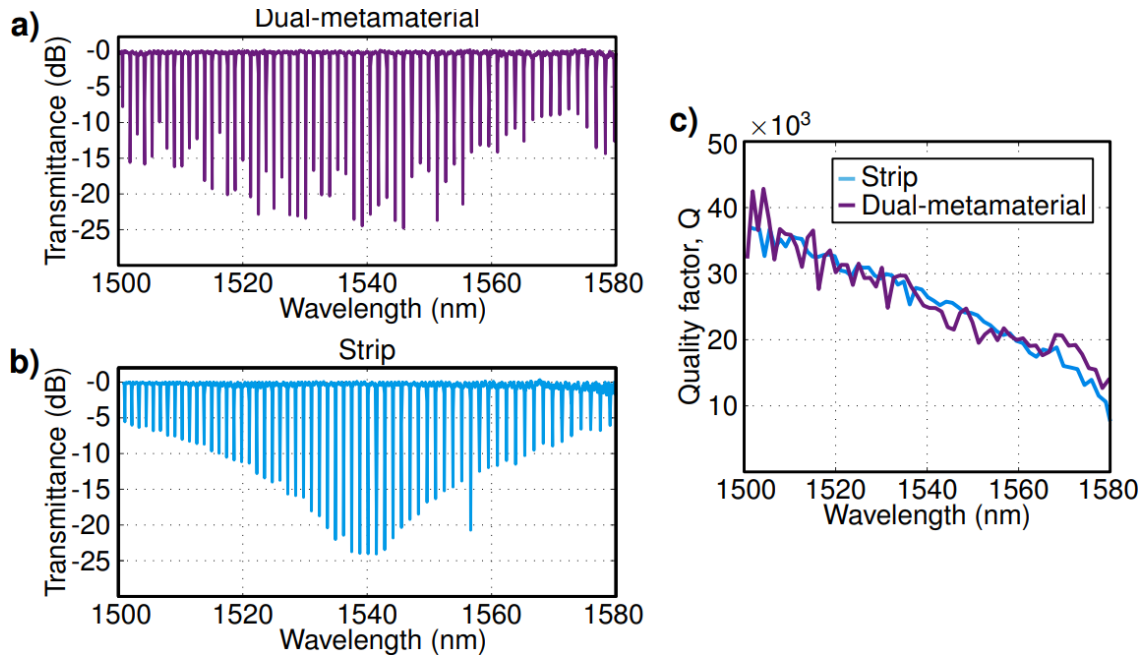

*Figure 6:* Comparison of the spectral response of dual-metamaterial and strip ring resonators, both at critical coupling with radius of 80 µm: a) Measured transmittance of nominal dual-metamaterial ring resonator ($\Delta$ = 0 nm) with a coupling gap of 600 nm. c) Measured transmittance for strip ring resonator with a coupling gap of 660 nm. c) Comparison of the measured quality factor as a function of the wavelength.

These results show that the proposed dual-metamaterial waveguide presents no penalty in terms of the quality factor, compared to strip waveguide. Comparing quality factors with other works may be somehow misleading as the performance of the resonators depends on the waveguide geometry and the specific fabrication technology, e.g. through the level of side-wall roughness. To the best of our knowledge, there is no demonstration of subwavelength nanostructured ring resonators in SOI with air cladding, as the high cladding asymmetry hampers the implementation of such resonators. Nevertheless, as a qualitative reference, previously reported micro-ring resonators with subwavelength nanostructures showed quality factors on the order of 10000 for liquid cladding (n~1.33) [3, 8]. Reported quality factors for slot rings with liquid cladding are in the 10,000-30,000 range [24].

## 4. Discussion and summary

We have proposed and experimentally demonstrated a new kind of nanophotonic waveguide that allows flexible control of the modal field confinement by implementing two different metamaterials in the core and the cladding. This dual-metamaterial geometry provides new degrees of freedom to shape vertical and horizontal index contrast, independently, and to engineer the single-mode condition. The metamaterial index in the core determines the vertical index contrast, while the difference between the metamaterials in the core and cladding sets the horizontal index contrast and the single-mode conditions. The indices in the core and cladding can be controlled with high precision and flexibility, just by judicious design of the holes in the subwavelength lattice, yet relying on a single etch step. The proposed waveguide enables the seamless implementation of wider single-mode waveguides with optimized vertical and horizontal evanescent fields that maximize overlap with the surroundings while overcoming limitations due to leakage into the substrate. To illustrate the potential of this approach we implemented dual-metamaterial ring resonators operating with TM polarization in SOI wafers with a 220-nm-thick guiding silicon layer, 3-µm-thick BOX, and air as a cladding. Such thin silicon guiding layer and asymmetric cladding hamper the implementation of slot and conventional SWG metamaterial waveguides due to strong propagation loss arising from leakage into the silicon substrate. Thus, strip waveguides are the preferred solution for applications where air serves as upper cladding, e.g. in gas sensing [18]. On the other hand, recent theoretical studies suggest that despite having better external confinement factors, slot and SWG metamaterial waveguides operating with TE polarization yield an inferior sensing performance than strip waveguides due to higher propagation loss, regardless of the cladding material [22]. Our 3D-FDTD calculations show that the TM mode of our dual-metamaterial waveguide yields better external confinement factors in the air (0.35) than strip waveguides with TM polarization (0.3). Still, the vertical confinement of the dual-metamaterial waveguide is similar to that of the strip waveguide, overcoming substrate leakage loss limitation of slot and SWG waveguides. Our experimental results show that strip and dual-metamaterial ring resonators fabricated with the same technology yield similar quality factors, near 30,000. Having similar quality factors and better external confinement factor, the proposed dual-metamaterial metamaterial ring resonators stand as a promising alternative to conventional silicon strip waveguides, to control modal confinement in silicon waveguides.

The dual-metamaterial approach proposed here opens a whole new range of possibilities to engineer the index profile in silicon waveguides no feasible in the state-of-the-art strip, slot or SWG metamaterial waveguides. For example, the dual-metamaterial waveguide enables the implementation of asymmetric index contrast configurations with different cladding metamaterials at each side of the waveguide, or graded index contrast implementations with effective metamaterial index gradually changing between the core and the cladding. These arrangements could be exploited for example to further relax single-mode condition. Furthermore, different configurations could be easily combined within the same silicon circuit requiring only a single etch step. The flexible index contrast engineering enabled by the dual-metamaterial structures can be instrumental for applications requiring strong evanescent fields, like biochemical sensing, but could also be key for advanced dispersion and confinement engineering in nonlinear applications [5]. We believe that these results open a fundamentally new route for the implementation of advanced modal confinement and dispersion engineering in silicon photonic devices.

## Acknowledgements

Agence Nationale de la Recherche (ANR) (MIRSPEC ANR17-CE09-0041, BRIGHT ANR-18-CE24-0023-01)